\documentclass[preprint2]{aastex631}

\usepackage[english]{babel}
\usepackage{natbib}
\usepackage{amsmath}

\def\ergsc{\hbox{ergs cm$^{-3}$ s$^{-1}$}}

\defcitealias{2016Beloborodov}{BL16}
\defcitealias{2015Sinha}{Sinha15}

\def\lsim{\mathrel{\rlap{\lower 4pt \hbox{\hskip 1pt $\sim$}}\raise 1pt \hbox
        {$<$}}}
\def\gsim{\mathrel{\rlap{\lower 4pt \hbox{\hskip 1pt $\sim$}}\raise 1pt \hbox
        {$>$}}}

\shorttitle{Magnetar Heating} \shortauthors{Tsuruta et al.}

\begin{document}

\righthead{Magnetar Heating}

\title{Ambipolar Heating of Magnetars}

\author{Sachiko Tsuruta}
\affiliation{Department of Physics, Montana State University, 
Bozeman, MT 59717, USA}
\affiliation{Kavli Institute for the Physics and Mathematics 
of the Universe (WPI), The University of Tokyo, Kashiwa, Chiba 
277-8583, Japan}
% ************************
%\\email: tsuruta@montana.edu}

\author{Madeline J. Kelly}
\affiliation{Department of Physics, Montana State University, 
Bozeman, MT 59717, USA}

\author{Ken'ichi Nomoto}
\affiliation{Kavli Institute for the Physics and Mathematics 
of the Universe (WPI), The University of Tokyo, Kashiwa, Chiba 
277-8583, Japan}
% ************************
%\\email: nomoto@astron.s.u-tokyo.ac.jp, mori@astron.s.u-tokyo.ac.jp}

\author{Kanji Mori}
%      *******
\affiliation{Research Institute of Stellar Explosive Phenomena, Fukuoka
University,\\8-19-1 Nakamura, Jonan-ku, Fukuoka-shi, Fukuoka 
814-0180, Japan}  
%\\email: mori@astron.s.u-tokyo.ac.jp}

\author{Marcus Teter}
\affiliation{Raytheon Technologies, 16510 E. Hughes Dr., 
Aurora CO, 80011, USA}
\affiliation{Department of Physics, Montana State University, 
Bozeman, MT 59717, USA}
%\\email: teter@montana.edu}

\author{Andrew C. Liebmann}
\affiliation{Department of Physics, Montana State University, 
Bozeman, MT 59717, USA}
%\\email: andrew.liebmann@montana.edu}

\accepted{by the Astrophysical Journal on February 9, 2023}

%%%%%%%%%%

\begin{abstract}
Magnetars, neutron stars thought to be with ultra-strong magnetic fields 
of $10^{14 - 15}$ G, are observed to be much hotter than ordinary 
pulsars with $\sim 10^{12}$ G, and additional heating sources are 
required. One possibility is heating by the ambipolar diffusion 
in the stellar core. This scenario is examined by calculating the models 
using the relativistic thermal evolutionary code without making the 
isothermal approximation. The results show that this scenario can be 
consistent with most of the observed magnetar temperature data.  
\end{abstract}

\section{Introduction} \label{sec:intro}

The soft gamma-ray repeaters (SGR) and the anomalous X-ray pulsars 
(AXP) are now considered to be magnetars, the same population of the 
ultra-strongly magnetized neutron stars with magnetic fields of the 
order of 10$^{14}$ - 10$^{15}$ G on the surface (e.g., 
\citealt{2008Mereghetti,2001Thompson,1998Heyl,2015Potekhin}). 
Activities in these stars are powered by the dissipation of strong 
magnetic energy (e.g., \citealt{2001Thompson}). Magnetars generally 
undergo long quiescent periods with persistent X-ray emission between 
a shorter recurrent phase of gamma-ray bursts (e.g., 
\citealt{2008Mereghetti}).  
			
\begin{figure*}%[t]  % ---------------------------------------------- Fig~1
\plotone{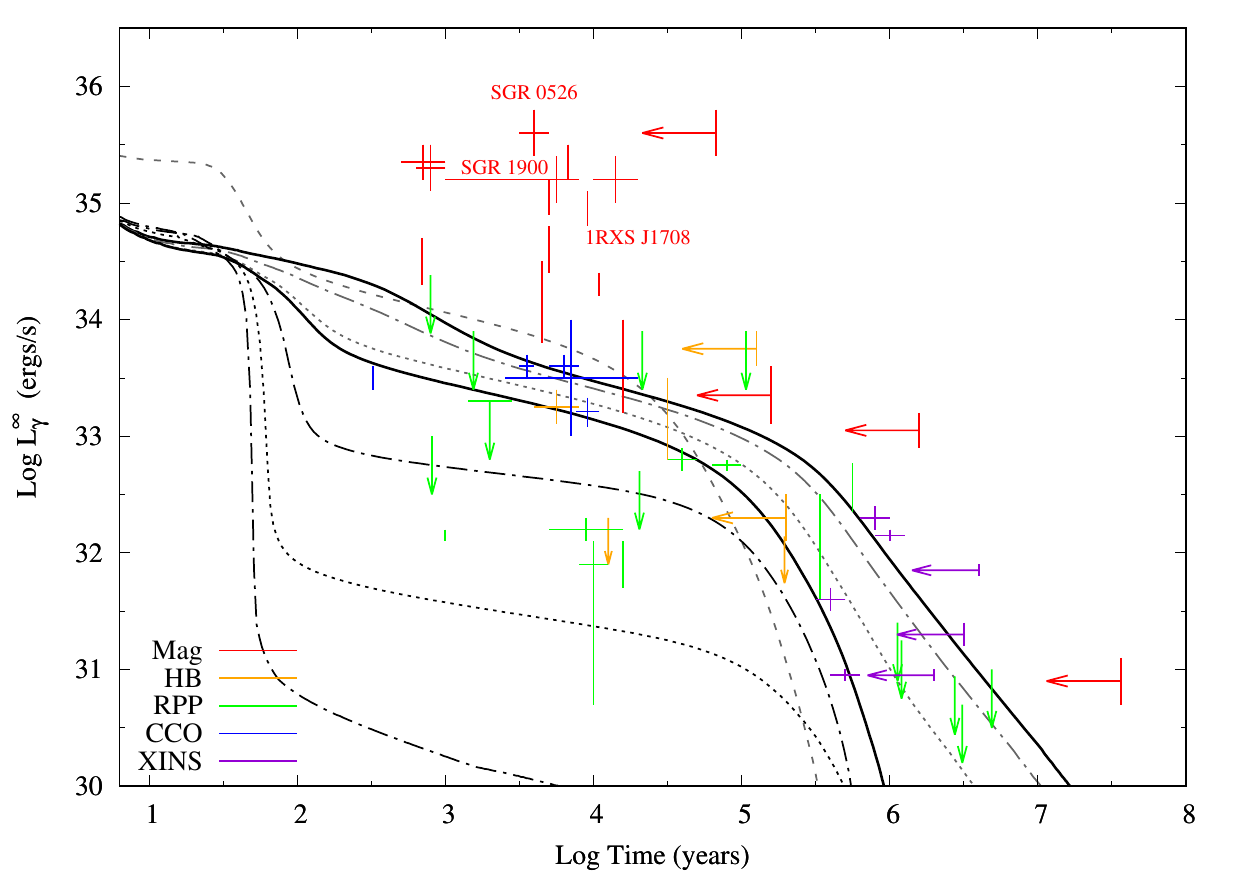}
\caption{Theoretical thermal evolution curves for various 
ordinary pulsars are compared with the observed surface photon 
luminosity (hence surface temperature) data of various kind of neutrons 
stars taken from \citet{2013Vigano}. The observed data points are 
grouped into five categories. Red are Magnetars (Mag), orange are 
High-B pulsars (HB), green are Rotation-Powered Pulsars (RPP), 
blue are Central Compact Objects (CCO) and purple are X-Ray 
Isolated Neutron Stars (XINS). The bars and crosses are detections.
Arrows pointing down represent upper limits on measured luminosity, 
arrows pointing to the left represent upper limits on age.  
See \citet{2013Vigano} for more complete details.} \label{fig:cool_curv}
\end{figure*}

During the quiescent phase the star is in a nearly steady equilibrium 
state. The surface temperature of many magnetars during this phase has 
been measured. Figure \ref{fig:cool_curv} shows the measured surface 
luminosity (and hence surface temperature) of magnetars (taken from
\citealt{2013Vigano}), which is compared with theoretical thermal 
evolution curves for ordinary neutron stars with magnetic fields of 
$10^{12}$ G (\citealt{2009Tsuruta_a}). 

%old figure 1 location

In this figure, two upper thick solid curves are for $1.4M_\odot$
neutron stars, with the lower curve for cooling only while the upper 
one includes the maximum vortex creep heating. Two curves between 
these thick solid curves show the vortex creep heating with intermediate 
strength. The hot dashed curve is for stars with the crusts 
contaminated by light elements. The lower three curves represent 
stars with $1.5M_\odot$, $1.6M_\odot$ and $1.8M_\odot$, respectively, 
in the order of decreasing luminosity. When the stellar mass 
reaches $M_{tr} = 1.45M_\odot$, the transition from neutron matter 
to hyperon-mixed matter takes place. Therefore, these are hyperon 
stars with the non-standard fast cooling. For the intermediate case 
of the $1.5M_\odot$ and $1.6M_\odot$ stars, superfluid suppression 
is effective. However, for the heaviest $1.8M_\odot$ star,  the 
central density is so high that the corresponding critical 
temperature for superfluidity becomes so low that the superfluid 
suppression disappears.

It is clear that most magnetars are hotter than ordinary neutron stars.
The surface temperature becomes higher with stronger magnetic fields, 
but with cooling alone the temperature can not become as high as the 
observed magnetar temperatures when the surface magnetic fields are 
increased even to as high as $10^{15}$ G (assuming the conventional 
magnetic field structure which becomes dipolar globally) 
(e.g., \citealt{1998Heyl}). With a certain special magnetic field 
configuration where ultra-strong magnetic fields of as high as 
$\sim 10^{16}$ G are deposited in the equatorial plane near the inner 
edge of the inner crusts, the star was shown to become as hot as the 
observed magnetars \citep{2013Vigano,2015Potekhin}. However, it was 
since then pointed out that this special magnetic structure is 
unlikely to last long due to instability (\citealt{2016Beloborodov},
hereafter\citetalias{2016Beloborodov}). Therefore, it is important 
to explore heating of magnetars with more conventional magnetic field
structure. 

In the following we only consider such a conventional magnetosphere case. 
There are several possible heating scenarios proposed for magnetars (e.g., 
\citetalias{2016Beloborodov}; \citealt{2001Thompson}). The star can be 
heated in the interior, either within the stellar core or in the crustal
layers. It can be heated from the outside from the magnetosphere also 
(see \citealt{2001Thompson},\citetalias{2016Beloborodov}).

\cite{2006Kaminker} considered the crustal heating of magnetars by 
using their thermal evolution code and explored the effect of the 
location of the heat source in the crusts and heating rate to the stellar
surface temperature. It was concluded that a heating rate of $\sim 3 \times 
10^{19}$ \ergsc~ is required at depths of 300 m or less, to sustain the 
surface radiation luminosity of $10^{35}$ ergs s$^{-1}$ required for
magnetars. These 
authors, however, did not consider the physical mechanisms for the 
heating, although some comments were made.

Some other authors considered various heating mechanisms for magnetars 
(e.g., \citealt{2001Thompson}, \citetalias{2016Beloborodov}). Recently 
\citetalias{2016Beloborodov} 
showed that a variety of heating mechanisms possible in the stellar 
crusts, such as the Hall effect, Ohmic dissipation, etc., will fail 
to increase the surface temperature to as high as the observed magnetar 
data.

Recently as one of the possible mechanisms for heating magnetars, 
\citetalias{2016Beloborodov} proposed the ambipolar 
process which takes place in magnetar’s central core. In this process 
magnetic energy is dissipated by the ambipolar drift which is the 
motion of the electron-proton plasma through the approximately 
static neutron fluid in the stellar core. The drift is driven by 
the Lorentz force. The ambipolar diffusion driven by the Lorentz 
force is opposed by proton-neutron friction and pressure gradients. 
These authors, however, adopted a simple analytic approach with a 
Newtonian model for a constant temperature and density core using the 
isothermal approximation.

Under the isothermal approximation, the core from the center to a
certain density $\rho_{\rm b}$ is isothermal where the timescale of
thermal conduction is assumed to be negligible
\citep{1964Tsuruta,1979Tsuruta}.
\citetalias{2016Beloborodov} adopted $\rho_{\rm b} = 10^{9}$ g 
cm$^{-3}$. Then the cooling and heating
effects of each layer are {\sl instantaneously} transported
throughout the core.  The thin outer envelope at $\rho < \rho_{\rm
b}$ has a spacially constant luminosity.  With this isothermal
approximation, the surface luminosity follows instantaneously the
change in the core temperature.  However, in the early stage of the
thermal evolution of the neutron star, the neutrino emission in the
core is much faster than the thermal conduction so that the surface
luminosity does not necessarily keep pace with the thermal evolution
of the core \citep{1981Nomoto, 1987Nomoto}.

We will, therefore, investigate this internal heating due to the 
ambipolar process, by fully taking into account the finite timescale 
of heat conduction in the core. We use the magnetar evolutionary 
simulation code which fully includes general relativity 
\citep{1977Thorne,1981Nomoto, 1987Nomoto} and realistic stellar 
physics with relevant equation of state (EOS).

In Section \ref{sec:mag_heating} we review the ambipolar diffusion  
process and then show how that will heat the stellar interior by the 
magnetic energy dissipated by this process. Section 
\ref{sec:model} introduces our physical model and summarizes our 
method and approach. Section \ref{sec:results} presents the results. 
The discussion and concluding remarks are given in sections 
\ref{sec:discussion} and \ref{sec:conclusion}. 

\section{Magnetar Heating by Ambipolar Diffusion} \label{sec:mag_heating}

\subsection{Ambipolar Diffusion Process} \label{subsec:diffusion_process}

The main process capable of dissipating magnetic energy in magnetar's 
core is diffusion through ambipolar drift 
(\citealt{1992Goldreich,1996Thompson}).  Ambipolar drift is the motion
of the electron-positron plasma through the (approximately static)
neutron fluid. The drift is driven by the Lorentz force 
  $J \times B/c = (\nabla \times B) \times B/4\pi$ and tends to 
relieve the magnetic stresses that drive it. The drift is opposed by (i)
friction against the neutron fluid and (ii) pressure 
perturbations it induces. The friction is due to nuclear collisions 
between neutrons and protons. (Electron-neutron collisions are negligible.)  

The rate of proton-neutron (p-n) collisions per proton, $\tau_{\rm pn}$, 
is  (\citetalias{2016Beloborodov})
\begin{equation} \label{eq:collision_rate}
%Equation 1
    \tau_{\rm pn}^{-1} \approx 5 \times 10^{18} T_9^2 (\rho/\rho_{\rm
    nuc})^{-1/3} Q_{\rm pn} \  {\rm s}^{-1},
\end{equation}\noindent
where $T_9$ is the temperature in the unit of 10$^9$ K, $\rho$ is the
density, and $\rho_{\rm nuc} \approx 2.8 \times 10^{14}$ g cm$^{-3}$ 
is the nuclear density. $Q_{\rm pn}$ describes suppression of the 
rate of collisions among protons and neutrons. It refers to proton 
(p) superconductivity and neutron (n) superfluidity. If there exist 
no proton superconductivity 
and no neutron superfluidity $Q_{\rm pn} = 1$. If  they are present, 
$Q_{\rm pn} < 1$ (see sections \ref{subsec:neutron superfluidity} and 
\ref{subsec:proton superconductivity} for the details.).

Pressure perturbations are induced if $\nabla\cdot(n_e {\bf v}) 
\not= 0$, where $n_e = n_{\rm p}$ are the electron (e) and proton 
(p) number density and ${\bf v}$ is the proton drift velocity. This 
compressive drift generates a change in $n_e$ which changes the 
electron and proton pressures. That is related to chemical potentials 
of electron and proton $\mu_e$ and $\mu_{\rm p}$. Then the resultant 
pressure gradient is given as $- n_e\nabla(\Delta \mu)$ where
\begin{equation} \label{eq:chemical_potentails}
%Equation 2
    \Delta \mu = \mu_e + \mu_{\rm p} - \mu_{\rm n}.
\end{equation}\noindent
$\mu_{\rm n}$, $\mu_{\rm p}$ and $\mu_e$ are, respectively, neutron,
proton and electron 
chemical potential. Equation \ref{eq:chemical_potentails} describes a local 
deviation from chemical $\beta$-equilibrium e,p $\longleftrightarrow 
n^2$. The chemical potentials include the rest mass of the species. 

The ambipolar diffusion is then given as (\citealt{1992Goldreich})

\begin{equation} \label{eq:ambipolar_diffusion}
%Equation 3
(\nabla \times B) \times B/4\pi = n_e \nabla(\Delta \mu) + 
n_e m_{\rm p}^\star {\bf v}/\tau_{\rm pn},
\end{equation}

\noindent
where $B$ is the core magnetic field and $m_{\rm p}^\star$ is proton
effective mass. The reaction rate is written as 
$dn_e/dt = - \vert \lambda \Delta 
\mu \vert$, where $\lambda$ is related to the compressibility 
of the plasma and given by 

\noindent
\begin{equation} \label{eq:plasma_compress} \begin{split}
%Equation 4
    \lambda \approx 5 \times 10^{33} T_9^6 (\rho/ \rho_{\rm nuc})^{2/3} 
%                                                 nuc==>*******
    H Q_\lambda \ \\
    {\rm ergs}^{-1} \ {\rm cm}^{-3} \ {\rm s}^{-1}.
\end{split}\end{equation}

\noindent
$Q_\lambda $ is suppression of $\lambda$ due to neutron 
superfluidity. $H$ refers to the enhancement due to the deviation 
from $\beta$ equilibrium.

Two regimes are possible: a friction dominated regime where $l \gg a$ 
and a pressure pillow regime with $l \ll a$, where
$l$ is a characteristic scale of the field variation 
$\Delta B$, and $a$ is a characteristic length defined and given by 
\citet{1992Goldreich} as
\begin{equation} \label{eq:char_lenght}
%Equation 5
    a = (\tau_{\rm pn} n_e/\lambda m_{\rm p}^\star)^{1/2}.
\end{equation}

Further details are found in \citetalias{2016Beloborodov}.

\subsection{Magnetar Heating by Ambipolar Diffusion} 
\label{subsec:abipolar_heating}

To calculate the ambipolar heating, we used (\citetalias{2016Beloborodov}) 
\begin{subequations} \label{eq:heating_equ}
%Equations 6
\begin{align}
dq_{\rm h}/dt \approx - B_1^2 b (dl_1/dt) /12\pi^2 \label{eq:mag_heat}
\break
\intertext{where}
dl_1/dt \approx - \tau_{\rm pn} B_1^2 /(2\pi \rho_{\rm p} l_1) \quad 
\text{for  } l_1 
\geq l_\star, \label{eq:amb_heating_region_one} \\
dl_1/dt \approx - \lambda B_1^2 l_1/(4\pi n_e^2) \quad \text{for  } 
l_1 \leq l_\star .\label{eq:amb_heating_region_two}
\end{align}
\end{subequations}

\noindent 
Here $dq_{\rm h}/dt$ is ambipolar heating rate, $l_1$ is the 
characteristic size of the 
ambipolar heating region. $l_\star = l_1$ when $l_1 = a\sqrt{2}$.
$B_1 = 2 B_0/\pi$, where $B_0$ is the peak magnetic field in the core. 
Initially $B = B_0$ sin $b x$, and $l_1$ and $b$ are related by 
$l_1 = 2/\pi b$. 

Sec 3.4 and Appendix in \citetalias{2016Beloborodov} give 
the details of their 1D 
(dimensional) model with the initial $B = B_0$ sin $bx$. Our 
simulation code is also 1D in the r direction. This (1D in the r 
direction) approach is essentially adopted in the `exact’ thermal 
evolution code by the experts of this trade, since the variables 
and parameters change mostly in the r direction, not in the angular 
directions. If the magnetosphere is radial the r direction flux in 
1D is just multiplied by the whole surface area to obtain the total 
photon luminosity. In our current paper the dipole magnetosphere is 
adopted. Then, by integrating the 1D flux in each direction 
including the angular dependence of the dipolar field geometry, 
it was found that the net effect of the geometry is that the radial 
geometry case with the polar direction flux will be reduced by about 
1/3. That is what our code does in this paper, to introduce the 
effect of magnetosphere geometry. For the 1D flux case we adopted 
the \citetalias{2016Beloborodov} model explained in detail in that paper.

\section{Magnetar Thermal Evolution Model} \label{sec:model}

In our magnetar model, heating by the ambipolar diffusion under
ultra-strong magnetic fields takes place in the central stellar core, 
while such ultra-strong magnetic fields also seriously alter the 
structure and property of the surrounding crustal layers. By taking 
into account these new features, we constructed the thermal 
evolutionary simulation code for magnetars, revising the latest 
version of our exact evolutionary code for ordinary neutron stars.

Our neutron star {\sl evolutionary} code has been developed over 
years - for the details, see
\citet{1981Nomoto,1987Nomoto,1986Tsuruta,1998Tsuruta,
2009Tsuruta_b,2009Tsuruta_a}. The set of general relativistic 
basic stellar structure evolution equations developed by 
\citet{1977Thorne} are solved simultaneously from the stelar 
center to the surface without making isothermal approximations. 
The evolutionary simulation code originally started by 
\cite{1981Nomoto, 1987Nomoto}, which has been continuously updated 
by adopting the most up-to-date microphysical input. See, e.g.,
\cite{1998Tsuruta}, \cite{2009Tsuruta_a}, and \cite{2018Tsuruta} 
for the details.

In the magnetar thermal evolutionary code constructed from the latest
ordinary neutron star evolutionary code, the neutrino emissivity 
consists of all possible standard mechanisms, with the modified Urca, 
plasmon, pair neutrino, photoneutrino, bremsstrauling, etc. The 
neutron superfluidity model with the critical temperature of 
log($T_{\rm crit}$(K)) = 9.45  (e.g., \citealt{2009Tsuruta_a}) is 
adopted. For the Cooper pair breaking and formation contribution, 
we found that many earlier publications, such as \cite{1999Yakovlev}, 
give only incomplete treatments. Therefore, we took the more recent, 
more up-dated version by \cite{2008Kolomeitsev}. 
In addition, the ambipolar heating (\ref{eq:mag_heat}) is adopted. 
In doing so, for the hotter earlier period, Equation 
(\ref{eq:amb_heating_region_one}) was used, while, during the 
later cooler period, Equation(\ref{eq:amb_heating_region_two}) was 
adopted.

This magnetar evolutionary simulation code was used for the 
evolution of the core from the center to the outer boundary chosen 
at  $M_r = M_{\rm core}$ where the matter density $\rho$ is as low as  
$\sim 10^9$ g cm$^{-3}$.  (Here $M_r$ is the baryon mass interior to 
the radius $r$.) With this code, we obtain the evolutionary change in 
the core temperature $T_{\rm core}$ at the outer boundary of 
$M_r = M_{\rm core}$.

For the equation of state (EOS), the maximum mass of the neutron star
has been found to be at least $2M_\odot$ \citep{2010Demorest}. 
\citet*{2008Takatsuka} constructed an advanced EOS model, which we 
refer to as HP8u, with the maximum mass going beyond $2M_\odot$. This model 
is based on the universal many body interactions among both nucleons 
and hyperons (see \citealt*{2008Takatsuka,2018Tsuruta} for the details). 
We used the HP8u EOS in our current work.  For this EOS, the core of lower 
mass, hotter stars consists mainly of neutrons and protons, while 
the major composition transforms to hyperons in heavier, cooler stars.
The mass of the star at this transformation point is $1.45M_\odot$. 
Since magnetars should be hot, in this paper we adopt  
hotter, less massive stars. Using the medium EOS HP8u (see,
e.g., \citealt{2018Tsuruta} for the details of this EOS model),
we constructed a neutron star model with $M = 1.4M_\odot$, 
where $M$ is the total baryon mass of the neutron star 
(see Section \ref{subsec:stellar model} for the details).
For this star, the central density is 1.0 $\times 10^{15}$ g cm$^{-3}$,
and the radius is $R = 11.7$ km. We adopt the core mass, 
$M_{\rm core}$, for this model at $1 - M_{\rm core}/M = 4.7 \times 
10^{-8}$, where $\rho_b = 2.0 \times 10^9$ g cm$^{-3}$.

For the envelope at $M_{\rm core} \le M_r \le M$, 
the spatially constant luminosity is a good approximation 
because of low densities. For the magnetar envelope, the 
ultra-strong magnetic fields significantly increase thermal 
conductivity in the crustal layers, which results in significantly 
reduced difference between the core temperature and the surface 
temperature \citep{2015Potekhin}. \citetalias{2016Beloborodov} 
calculated such magnetar envelope models and showed the results 
in their Figure 1. We approximate their relation between the core 
temperature, $T_{\rm core}$, and the surface photon luminosity,
$L^{\infty}_{\gamma}$, and obtained the following equations for 
four set of $B$ ($3 \times 10^{13}$ and $10^{15}$ G) and chemical 
composition (heavy element dominated such as Fe and light element
contaminated).

 \begin{subequations} 
 \noindent
For the $B = 10^{15}$ G and Fe case:
\begin{align} \label{eq:core_lum_equ_a} \begin{split}
%Equation 7a
 \shoveleft   
 {\rm log}_{10}[L^{\infty}_{\gamma}({\rm ergs~s}^{-1})] = \\
    1.96 \times 
    {\rm log}_{10} [T_{\rm core}({\rm K})]  + 17.446
       \end{split} \end{align}
For the $B = 3 \times 10^{13}$ G and Fe case:
        \begin{align} \begin{split}
        {\rm log}_{10}[L^{\infty}_{\gamma}({\rm ergs~s}^{-1})] = \\
        2.03 \times {\rm log}_{10} [T_{\rm core}({\rm K})]  + 16.583
         \end{split} \end{align}
For the $B = 10^{15}$ G and light element case: 
\begin{align} \begin{split}
    {\rm log}_{10}[L^{\infty}_{\gamma}({\rm ergs~s}^{-1})] = \\ 1.50 
    \times {\rm log}_{10} [T_{\rm core}({\rm K})]  + 21.869
%equaton 7c
\end{split} \end{align}

For the $B = 3 \times 10^{13}$ G and light element case:
\begin{align} \label{eq:core_lum_equ_d} \begin{split}
    {\rm log}_{10}[L^{\infty}_{\gamma}({\rm ergs~s}^{-1})] = \\ 1.50 
    \times {\rm log}_{10} [T_{\rm core}({\rm K})] + 21.673
%equation 7d
\end{split} \end{align}
\end{subequations}

These equations are good approximation of Figure 1 of
\citetalias{2016Beloborodov}.

\section{Results} \label{sec:results}

Using the magnetar thermal evolutionary code as described in 
the previous section, the thermal evolution of six representative 
cases are calculated. The properties of these cases are shown in 
Table \ref{tab:B_field_kappa_table}. $B$ is the core magnetic field, 
and $b$ is a parameter defined in Section \ref{sec:mag_heating}. 
As a representative case we chose $B$ to be $\sim 10^{16}$ G, because 
the earlier work (e.g., \citealt{1998Heyl}) shows that for magnetars 
the core magnetic field will be about 10 times the surface magnetic 
field which is estimated to be about $10^{15}$ G. The first three 
are the cases adopted by \citetalias{2016Beloborodov}, while the 
last three are the additional choices with the different relevant 
combinations of $B$ and $b$ within the acceptable range.    

\begin{figure*}[t]  % ---------------------------------------------- Fig~2
\plotone{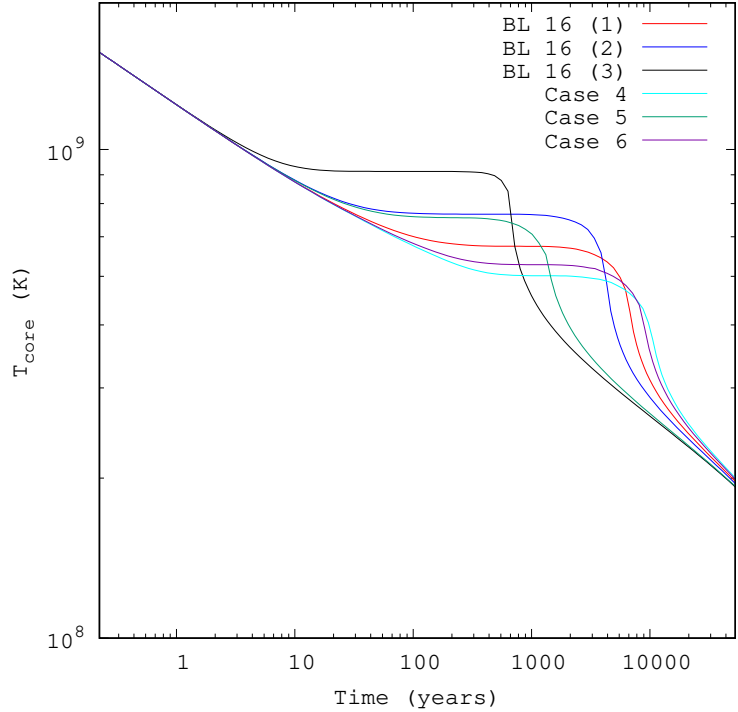}
\caption{The core temperature vs time from our simulations, for six 
representative cases in Table \ref{tab:B_field_kappa_table}. The 
first three cases are the same as in \citetalias{2016Beloborodov}, 
to facilitate comparison with the \citetalias{2016Beloborodov} 
curves. The last three choices are of our own. See Table
\ref{tab:B_field_kappa_table}  for the details of the cases.}
\label{fig:core_temp_ours}
\end{figure*}

The results for the six representative cases (see Table
\ref{tab:B_field_kappa_table}) are shown in Figure 
\ref{fig:core_temp_ours}, where the core temperature vs. age is shown. 
During the earliest stages the core temperature is too high 
($T_{\rm core}$ $\gsim$ $10^9$ K), for the heating to effectively  
compete with the neutrino cooling, and the star cools essentially by the
escaping neutrinos. However, the heating becomes sufficient to balance 
cooling as the star cools down to around that temperature. Thereafter 
the curve follows the plateau region while cooling is balanced by heating. 
After the evolution reaches the characteristic time for the magnetic 
energy dissipation by this process (\citetalias{2016Beloborodov}), the 
curve goes down again, approaching the cooling curve.

\begin{table} 
\begin{tabular}{lcc}
Model Name & $B$ ($\times$10$^{16}$ G) & $b$ \\ \hline \hline
BL16 (1)   & 1      &  $\pi\times$10$^{-5}$ \\
BL16 (2)   &  1.5         & $\pi\times$10$^{-5}$ \\
BL16 (3)   &  1.5        &  10$^{-6}$ \\ \hline
Case 4     &   0.7        &   $\pi\times$10$^{-5}$ \\
Case 5     &    0.8       &   10$^{-6}$ \\
Case 6     &    0.8       &   $\pi\times$10$^{-5}$\\ \hline
\end{tabular}
\caption{Properties of the six representative cases used in 
Figure \ref{fig:core_temp_ours}. The core magnetic field strengths 
$B$ and the parameter $b$, as defined in Section   \ref{sec:mag_heating},
are shown for each case.}\label{tab:B_field_kappa_table}
\end{table}

%figure 2

The ultra-strong fields of magnetars also affect strongly the crustal 
layers from the radial position at $M_{\rm core}$ to the stellar surface, 
which alters significantly the relation between the core temperature and 
surface temperature. It depends on various microphysics of the crustal 
layers. 

The most important is the composition. For instance, under the ultra-strong 
magnetar fields thermal conductivity is much higher for light elements such 
as Hydrogen, as compared with heavy elements \citep{2007Potekhin}. The 
dominant heavy element is Fe. But since magnetars are still relatively 
young, it is quite reasonable that the crusts are still contaminated by 
light elements in some cases. Equations \ref{eq:core_lum_equ_a} to
\ref{eq:core_lum_equ_d} show the surface radiation as a function of core 
temperature for both cases, with two representative surface magnetic field 
strengths of $10^{15}$ G and $3 \times 10^{13}$ G.

\begin{figure*}%[t]  % ------------------------------------------------- Fig~3
\plotone{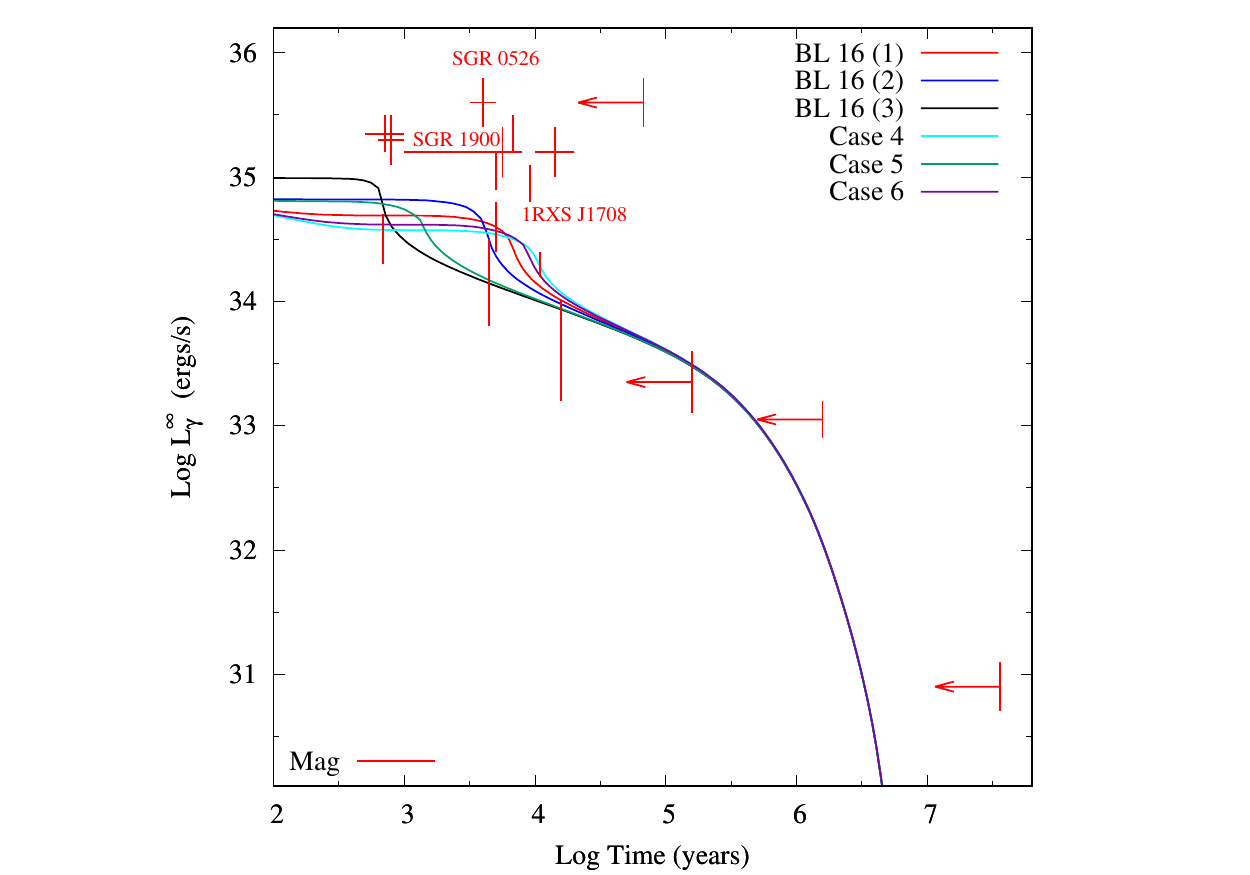}
\caption{ Thermal evolution of magnetars with the ambipolar 
heating is compared with magnetar observation data. The six cases in 
Table \ref{tab:B_field_kappa_table} for models with the major crustal 
composition of Fe, are shown as the surface photon luminosity vs age.  
The bars and crosses are detections. The horizontal arrows are 
the upper limit to the age. The theoretical curves are consistent 
with cooler magnetar data.} 
\label{fig:magnetar_cooling_cur_fe}
\end{figure*}

\begin{figure*}%[t]  % ----------------------------------------------- Fig~4
\plotone{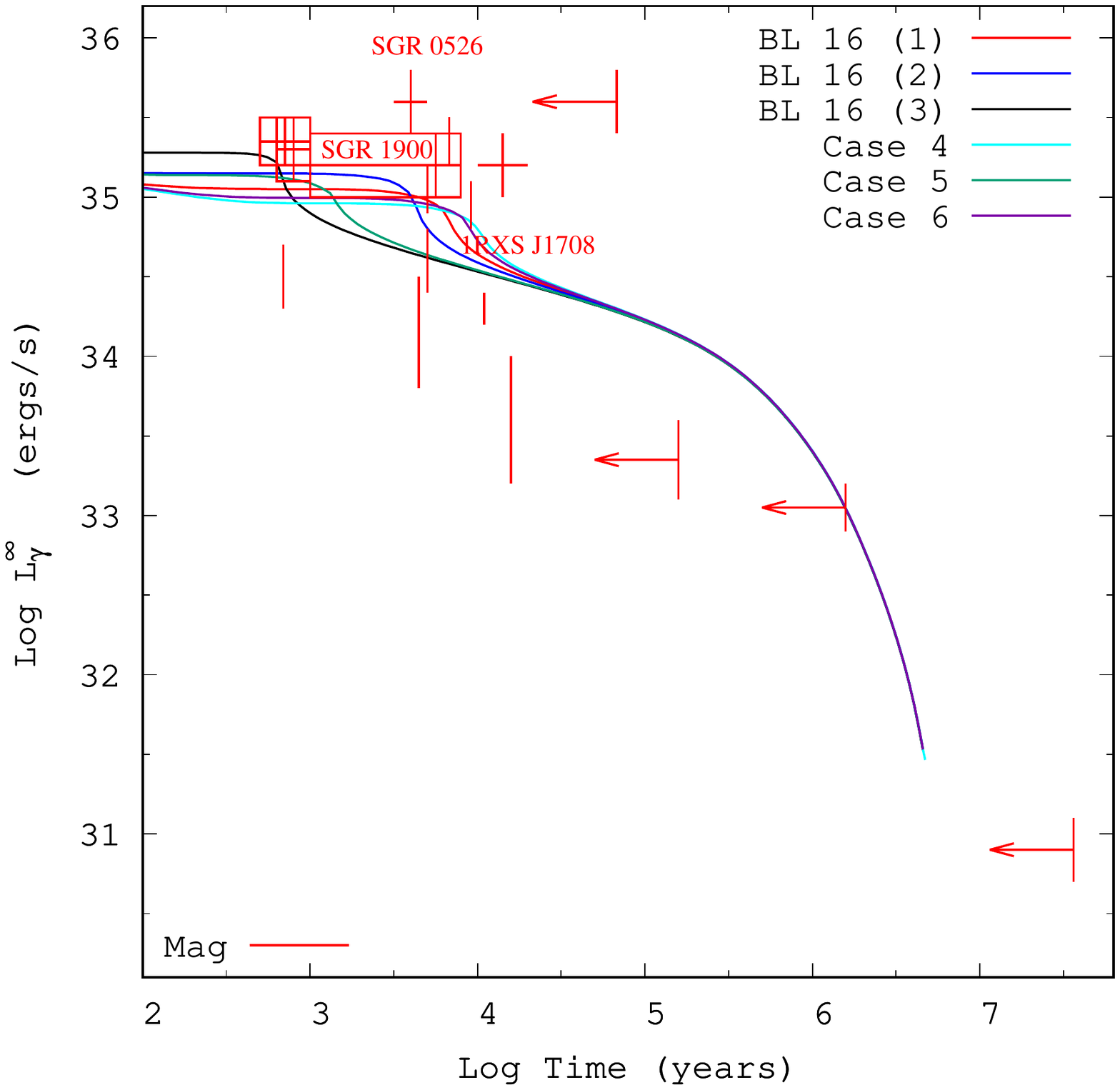}
\caption{The same as Figure \ref{fig:magnetar_cooling_cur_fe}, except 
that the crustal layers of the star are now contaminated by light 
elements. The theoretical curves are now consistent with many of hotter
magnetar data.  In this figure, to avoid overcrowding, only the 
$1\sigma$ errors are shown. When $3\sigma$ errors are included, 
the data points of only a few hottest sources are still off the 
curves.} 
\label{fig:magnetar_cooling_cur_he}
\end{figure*}

For the six representative cases in Table \ref{tab:B_field_kappa_table}, 
thermal evolution (cooling and heating) curves are shown, as the surface 
photon luminosity vs. age relation for the Fe envelope case in Figure 
\ref{fig:magnetar_cooling_cur_fe}, while the case for the light 
element contaminated envelope is shown in Figure 
\ref{fig:magnetar_cooling_cur_he}.
The surface magnetic field used is $10^{15}$ G. For comparison the
observed surface luminosity data of various magnetars with different ages,
taken from \citet{2013Vigano}, are also shown.

First consider the evolution of the central temperature. 
Our core temperature vs. age relation is shown in Figure 
\ref{fig:core_temp_ours} for the six representative cases in Table
\ref{tab:B_field_kappa_table}. Similar core temperature evolution 
for the isothermal model was obtained 
by \citetalias{2016Beloborodov} for the top three cases of our Table
\ref{tab:B_field_kappa_table}, and that is shown in their Figure 3. 
Comparing Figure 3 of \citetalias{2016Beloborodov} and our Figure
\ref{fig:core_temp_ours}, qualitatively it may appear that the effect 
of the ambipolar heating on the core temperature evolution is similar 
between their isothermal model and our model. However, 
some important difference is noted. 

For instance, in our case the curves are generally more smooth, 
especially during the decaying stages. In our model the transition 
from the earlier to the later period is smooth, not abrupt. More 
importantly the plateau and decaying phases in our case is more 
gradual and longer (compere their Figure 3 and  our Figure
\ref{fig:core_temp_ours}).  That makes the temperature {\it higher} 
in our model during the decaying phase. {\it That is important because 
that is the critical location of most observed temperature data}. 

Now we consider the thermal evolution behavior expressed as the 
evolution of the surface photon luminosity (and hence the surface 
temperature) vs. age. We calculated the surface temperature evolution 
for $B$ = $10^{15}$ G, and the results are shown in Figures
\ref{fig:magnetar_cooling_cur_fe} and \ref{fig:magnetar_cooling_cur_he}. 
Figure \ref{fig:magnetar_cooling_cur_he} shows that in the 
the light element case {\it the surface temperature becomes 
significantly higher.} Compare Figure \ref{fig:magnetar_cooling_cur_fe} 
and Figure \ref{fig:magnetar_cooling_cur_he}. Note that the Fe 
envelope case (Figure \ref{fig:magnetar_cooling_cur_fe}), where 
the temperature is lower, still agrees with cooler observed data, 
but not some hotter ones. On the other hand, in Figure 
\ref{fig:magnetar_cooling_cur_he} with the light element case, due 
to the higher surface temperature, theoretical curves cover more higher 
temperature observed data.  In this figure, to avoid overcrowding, 
only the $1\sigma$ errors are shown. When $3\sigma$ errors are included, 
the data points of only a few hottest sources are still off the curves.   

Here the major reasons are summarized: First of all, the star 
still cools with escaping neutrinos from the core for magnetars 
due to their relatively younger ages, and hence the evolution is 
determined by the core temperature. For the same core temperature,
the surface temperature is higher in the light element case than the 
case with only heavy elements, due to the enhanced crustal thermal 
conductivity for light elements with ultrastrong field magnetars.

In conclusion, combining Figure \ref{fig:magnetar_cooling_cur_fe} for 
the Fe crust case and Figure \ref{fig:magnetar_cooling_cur_he} 
for the light element contaminated case, the observed magnetar 
temperature data can mostly agree with the theoretical predictions 
from the ambipolar heating. 

\section{Discussion} \label{sec:discussion}

\subsection{Comparison with Earlier Work} \label{subsec:comparison}

\citetalias{2016Beloborodov} did not convert their core temperature 
evolution results (shown in their Figure 3) to the surface temperature 
evolution case. Instead, these authors converted the observed surface 
temperature data to the corresponding core temperature data 
in their Figure 3 which are shown as an green box. By comparing that 
with their core temperature evolution curves, it is noted that the 
observed temperature is too high for their ambipolar heating scenario, 
because the green box lies above their ambipolar heating curves. 
The reason is that when these authors converted the observed surface
temperature data to the core temperature data they did not include the 
light element contaminated crust case. If they had done so, the lower 
end of their green box would have gone down considerably, getting closer 
to their heating curves. Their abstract does note that if the light 
element case is included, the observed data are consistent with the 
ambipolar heating. By examining \citetalias{2016Beloborodov}'s Figure 
1 where the surface radiation vs. core temperature relations are shown, 
it is clear that for the same surface luminosity (and hence surface
temperature), the corresponding core temperature should be 
significantly {\it lower} for the light element case (the red curves)
than the Fe cases (blue curves) in their Figure 1. That means the 
bottom of the green box in their Figure 3 should be much lower when 
the light element case is included. 

Another significant difference is that \citetalias{2016Beloborodov} 
adopted the modified Urca (their Murca) alone, while our neutrino 
emission includes the effects of neutron superfluidity. The Cooper 
pair breaking and formation contribution, which we included, increases 
the neutrino emission immediately below the critical temperature 
$T_{\rm crit} = 10^{9.45}(K)$. However, this Cooper pair emission 
decreases rapidly to below the superfluid suppressed case, which is 
{\it below} the Murca alone case. By the time the core temperature 
reaches the typical temperatures of the magnetars, at around 
$3 \times 10^8$ K, the neutrino emission issignificantly {\it below} 
the Murca alone case. See, e.g., Figure \ref{fig:core_temp_ours} of 
\citetalias{2016Beloborodov}, where a similar neutron superfluid 
model (red curves) is shown. Due to our overall reduced neutrino 
emission at around the age of magnetars, our heating lasts longer.

Another reason for the difference between \citetalias{2016Beloborodov} 
and our results is that when we adopt our more realistic numerical 
evolutionary simulation code the decaying phase of the heating curve 
is more gradual, not abrupt which happens in the 
\citetalias{2016Beloborodov} case. 
(Compare their Figure 3 and our Figure \ref{fig:core_temp_ours}). 
That also results in the surface temperature of our model higher for 
longer periods during the decaying phase where most of the magnetar 
observational data are located. 

\subsection{Effects of Neutron Superfluidity} 
\label{subsec:neutron superfluidity}

$Q_{\rm pn}$ describes suppression of the rate of collisions 
among protons and neutrons. We set $Q_{pn} =1$ for the effect of 
neutron superfluidity on the ambipolar heating since numerical information
is not available. However, we investigated its qualitative net effect
to our results, and reached the following conclusion: Heating rate
$dq_h/dt$ in  Eq.(6a) is proportional to $dl_1/dt$, which goes as 
$\tau_{pn}$ (Eq. (6b)). $\tau_{pn}$ goes as $1/Q_{pn}$ (Eq.(1)). 
With superfluidity, $Q_{pn} < 1$. Then $\tau_{pn}$ is larger with 
superfluidity. Then, ambipolar heating $dq_h/dt$ gets higher because 
it goes as $dl_1/dt$ (Eq.(6a)), which goes as $\tau_{pn}$ (Eq.(6b)). 
That means the ambipolar heating will increase with neutron superfluidity. 
(Note that \citetalias{2016Beloborodov} also reached the same conclusion.)
Therefore, if neutron superfluid is included in our heating calculations, 
heating will increase. That means heating will last longer, and our 
final conclusion on the validity of the ambipolar heating will become 
even stronger.

\subsection{Effects of Proton Superconductivity} 
\label{subsec:proton superconductivity}

The effect of superconductivity should be considered if the 
field $B < H_{c2}$ (Sinha \& Sedrakian (2015), here referred to as 
\citetalias{2015Sinha}). For our neutron star model, the central density 
$\rho_c$ is $10^{15}$ gm cm$^{-3}$. Table 1 in \citetalias{2015Sinha} 
shows that for that density, $H_{c2} < 0.03 ( \sim 3 \times 10^{14}$G). 
Since our core $B \sim 10^{16}G >> H_{c2}$ in the central core, the 
effect of superconductivity should be negligible. Note that $H_{c2}$ 
depends on density. \citetalias{2016Beloborodov} also note that 
superconductivity is negligible for mangneters. (Near the core boundary,
the density drops to near the nuclear density and then 
$H_{c2} > 10^{16}$G (see \citetalias{2015Sinha}, Table 1), but neutron 
star core is almost constant density. Therefore the volume where the 
density drops to the nuclear density region near the core boundary 
is negligible, compared with the whole core volume. Therefore, the 
contribution from near the edge of the core is negligible.)

\subsection{Effects of Stellar Model} \label{subsec:stellar model}

When the stellar mass gets beyond $1.45M_{\odot}$, our EOS model 
includes hyperons. Our current paper is to test if a magnetar model 
CAN get hot enough to be consistent with the observed magnetar data 
with the ambipolar heating. We showed, for the low mass 
$1.4M_{\odot}$ star, that is yes. With the EOS we adopted (which is 
perfectly relevant), this happens to be still a neutron star. 
Hyperon-mixed stars are heavier and cooler, and with our EOS model 
it is harder to heat more massive and cooler hyperon-mixed stars 
enough to the level of many of the observed hotter magnetar data. 
We showed that at least for less massive stars (which in our model 
happens to have no hyperons), this high heating is possible. In some other 
relevant EOS models where transformation to hyperons takes place at 
lower density and mass (e.g., \citealt*{2018Raduta}), low mass 
stars may contain hyperons already. For these stars  the same 
conclusion applies. Whether hyperons are present or not is not the 
issue in our current particular problem. It shows that, at least 
for lower mass hotter stars, the ambipolar heating is consistent with 
most of the observed magnetar data.

\section{Conclusions} \label{sec:conclusion}

We investigated the possibility of ambipolar heating in the stellar core 
as the source of magnetar's high temperatures. Our results are: 

(i) If both the heavy element crust case and the light element 
contaminated crust case are considered, the ambipolar heating can be 
consistent with most of observed magnetar temperatures.

(ii) If neutrons are in the superfluid state, the ambipolar 
heating increases, and the heating
phase can last longer, to be consistent with the observed magnetar
temperatures see Section \ref{subsec:neutron superfluidity}).

(iii) The adoption of the relativistic thermal evolution simulation 
code with a realistic magnetar model results in the evolution becoming
more smooth and gradual. That helps higher temperatures during the
decaying phase.

By adopting an isothermal, non-relativistic model with an analytic 
approach, \citetalias{2016Beloborodov} predicted that, if the stellar
crust is contaminated by light elements and the core magnetic field 
can be as strong as $10^{16}$ G, the ambipolar heating may heat the 
magnetars to the observed temperatures. At the same time, these 
authors pointed out that such a high temperature phase may not last
long enough.

Our results mostly confirm their estimates. However, it may be 
pointed out that the magnetars may last long enough especially if the 
neutrons in the interior are in a superfluid state, which is quite 
possible. Also it is quite possible that the magnetic field in the 
interior can be as high as $10^{16}$ G (e.g., see 
\citealt{2015Potekhin, 1998Heyl}). Also it is quite reasonable that 
magnetars still contain light elements due to their relatively 
young age.

Our conclusion is that the ambipolar heating 
can be consistent with the observed high temperature of magnetars.

\bigskip
\noindent
ACKNOWLEDGEMENTS

\noindent
We thank the referee for constructive suggestions and comments. 
This work has been supported by the Kavli IPMU and the World 
Premier International Research Center Initiative (WPI), MEXT, Japan,
and the Japan Society for the Promotion of Science (JSPS) KAKENHI 
grants JP17K05382, JP20K04024, JP21H04499, and JP21K20369. 

\bibliographystyle{aasjournal}
\bibliography{nsbib}

\end{document}